# Protein Folding from the Perspective of Chaperone Action

Samuel Nkrumah

**ABSTRACT:** Predicting the three-dimensional (3D) functional structures of proteins remains an important computational milestone in molecular biology to be achieved. This feat is hinged on a clear understanding of the mechanism which proteins use to fold into their native structures. Since Levinthal's paradox, there has been a lot of progress in understanding this mechanism. Most of the earlier attempts were caught between assigning either hydrophobic interactions or hydrogen bonding as the dominant folding force. However, a consensus now seems to be emerging about hydrogen bonding being a stronger force. Interestingly, a view from chaperone action may further throw some light on the nature of the folding mechanism. Thus the very mechanisms which prevent protein aggregation and misfolding, could help us have a better understanding of the folding mechanism itself.

**INTRODUCTION:** Proteins are an important class of biological macromolecules which function in cells to support structure, and facilitate movement and communication in different cellular locations. They are made up of chains of amino acids which form their primary structure. After a protein's primary structure is synthesized by the ribosome, it must then fold into it's 3D structure to be functional. In protein biochemistry, structure informs function, and knowing the structures of the many proteins that have been sequenced, will help us tell a more vivid story of the biological processes happening in cells. Unfortunately, the experimental methods which have enabled us to know the structures of most proteins are slow at revealing these structures. As an alternative, fast computational methods can be used to predict protein structures on account of the computing power available to us today. Compared to experimental methods, accurate protein structure prediction will be an efficient approach to drug design. But to do this requires a thorough understanding of how proteins fold. Because, from their unfolded states, proteins have a large number of starting points and paths (conformations, in the order of $9^n$, where n is number of amino acids) they can use to get to their native states.[1] If proteins were to fold by try and error, it will take them an unreasonable amount of time to fold. However, they fold by a mechanism that leads to faster folding times.[19] This means that given the best computing power, we still cannot reasonably fold proteins on computers without a similarly powerful algorithm. And the best algorithm can only come from a clear understanding of the mechanism proteins use to fold. This mechanism which has engaged the curiosity of several minds is the focus of this review. The literature in this field of research covers two approaches to the protein folding mechanism: one where hydrophobicity is the major folding force and another with hydrogen bonding as the primary folding force.

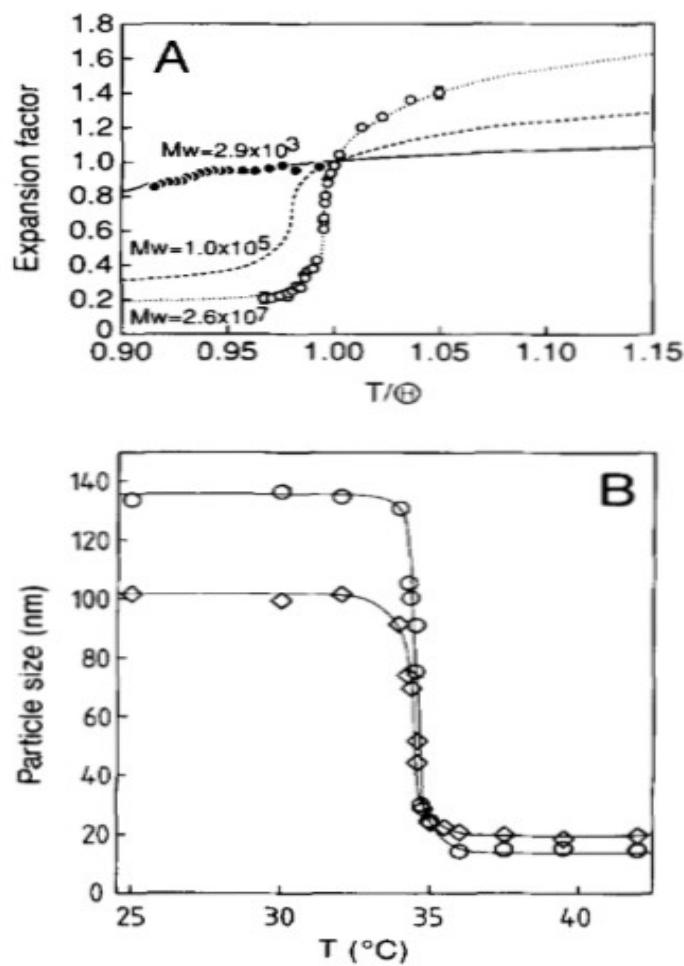

FIG 1. **(A) Hydrophobic collapse of polystyrene in cyclohexane**. **(B) Hydrodynamic radius and radius of gyration of poly-(N- isopropylacrylamide) in water.** Horizontal and vertical scale show temperature and polymer size respectively.[3]

***Hydrophobicity argument:*** The hydrophobicity basis of folding comes from the fact that non-polar substances are entropically favored to aggregate in water. Interestingly, studies **(Fig 1)** by Sun et al and Fujishige et al showed that collapse of long chain hydrophobic polymers is sharp in unfavorable solvents.[4,5] This led to the view that in long-chain proteins with more hydrophobic monomers, hydrophobic collapse could be a strong folding force. Indeed, off-lattice studies **(Fig 2)** by Yee et al showed that compaction induces secondary structure in polyalanine chains, although strict definition of secondary structure limited the amount of secondary structure observed.[3] And protein compactness was also found to increase amounts of secondary structure in DnaK and myoglobin.[6,7] Hydrophobicity thus seems like a good candidate folding force. Folding models such as the Collapse Model and Zipping and Assembly (ZAM) are based on this assumption.[3] ZAM for instance was used to predict the structures of seven out of nine proteins to <3Å rmsd from their experimental structures. However, it was found in a different study to be able to predict the structures of only a subset of proteins.[30] This raised concerns as to whether ZAM is a general folding mechanism.

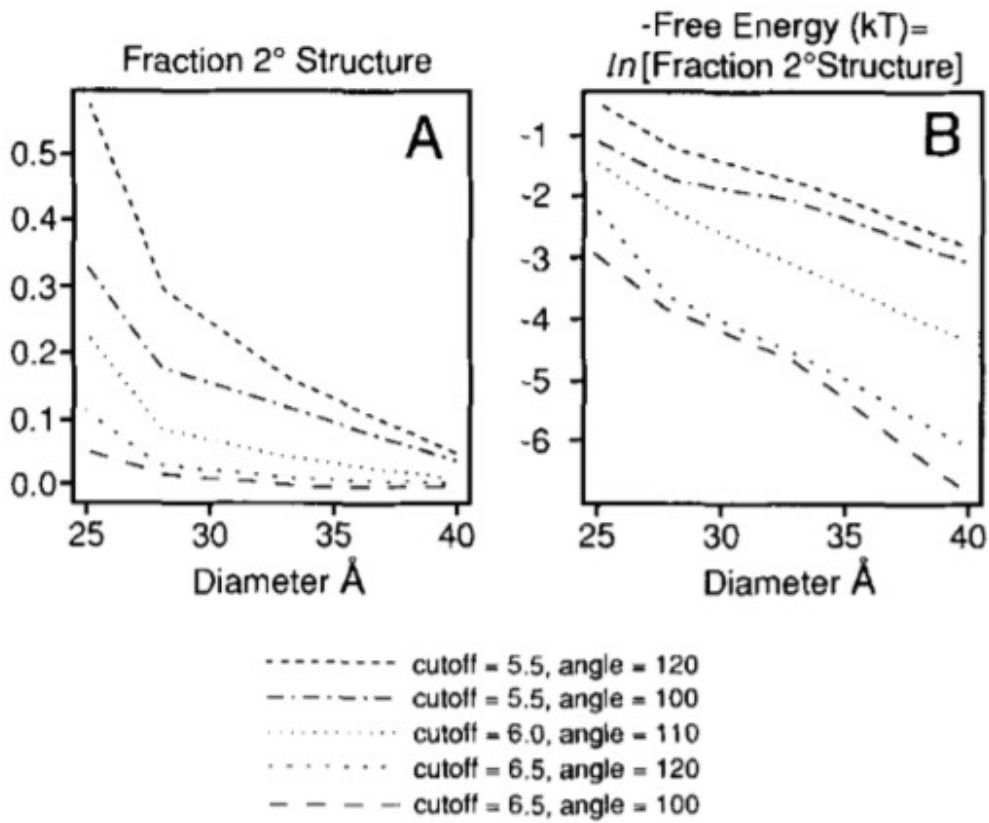

FIG 2. **(A) Amounts of secondary structure observed in off-latice studies by Yee et al depend on criteria used to define secondary structure.** Strict definition of secondary structure (Cutoff 5.0, angle=110) does not produce much secondary structure. **(B) Entropic stabilization due to compactness is independent of criteria used to define secondary structure.**[3]

*Hydrogen-bonding argument:* Ever since Pauling et al predicted the existence of hydrogen-bonded alpha-helices and beta sheets, hydrogen bonding has been considered the other candidate potent folding force. In their diffusion-collision model, Karplus and Weaver predict that proteins fold by first adopting secondary structures in microdomains (hydrogen bonded secondary structures), followed by coalescence of the microdomains into the final 3D native structure.[2] This model leads to a reduction in the area of conformational search space as well as overall short folding times. It is also consistent with the "new view" of folding which emerged predicting that folding can occur along multiple routes rather than a single pathway.[9,10] The main appeal of the diffusion-collision model is that, it reduces the complexity of the folding problem by breaking it into solvable pieces, sort of a divide and conquer approach. This same method is adopted by ZAM, although ZAM emphasizes hydrophobicity as the major force driving structure formation. A backbone folding theory hinged on hydrogen bonding has also been proposed by Rose et al.[11] They highlight several studies that provide compelling evidence for hydrogen bonding as the major folding force. According to the backbone theory, proteins, guided by backbone hydrogen bonding, fold in a hierarchical domain-wise fashion. It sees the commonality of protein backbone structure as a clue to a general backbone hydrogen bonding folding mechanism. Finally, Englander et al have shown through a combined hydrogen exchange and mass spectrometry technique that proteins fold by forming secondary structural elements in sequential steps that lead to the native structure.

In an interesting way, the field seems to be reaching a consensus on the major folding force. Recently, Ken Dill, who with colleagues proposed ZAM, has reported a new quantitative folding mechanism based on hydrogen bonded secondary structures.[12] As it is in science, any good model must be supported by experimental facts. And at the moment, there seems to be a lot of experimental data supporting a folding mechanism, with hydrogen bonding as the major folding force. The new model proposed by Dill has a lot of similarities with the diffusion-collision model. For instance, both models propose a microdomain-wise folding mechanism and predict folding to occur along multiple routes. This apparent consensus is a good thing. However, we need to take the folding mechanism from another point of view in order to get to the truth. For a good strategy for solving any problem is the consideration of different perspectives and multiple approaches.

**The Folding Mechanism from the view of Chaperones:** Chaperones are proteins which assist other proteins to fold into their correct native structures. In the crowded cytosol, they do so by preventing misfolding and hydrophobic aggregation of newly synthesized proteins. Insofar as chaperones promote protein folding, they can provide new insights or further corroborate existing theories on the folding mechanism. One such chaperone is the bacterial Trigger Factor (TF) which is known to associate with ribosomes and interact with nascent polypeptides in a co-translational manner.[13] From the mechanism of TF action on polypeptides, I present and discuss two theories about the folding mechanism.

**The Domain Folding Theory:** In newly synthesized polypeptides, TF has been reported to recognize basic and aromatic (hydrophobic) residues which occur on average every 32 residues.[16] *In vitro* studies have also demonstrated that multiple TF molecules can bind nascent chains, and that each TF molecule has a mean ribosome association time of ~10s, which corresponds to the synthesis of a protein domain.[17] It is also reported that each TF molecule after dissociating from the ribosome, can continue to interact with the chain for ~35s.[17] In this mechanism of TF, a first TF molecule binds to at least three hydrophobic residues on the first domain during protein synthesis. **(Fig 3)** This achieves the purpose of shielding the hydrophobic residues and preventing their aggregation early on in folding. After domain synthesis, TF dissociates from the ribosome still bound to the domain, and in the span of ~35s supports domain folding via hydrogen bonded secondary structures - which occurs in a microsecond-millisecond range – [19] stabilizing and releasing only after the next TF supported domain has folded in a similar fashion. This ripple of folding which travels along the protein chain may be catalyzed by aggregation of the released hydrophobic residues, which will lead to inter-domain stabilization and ultimately to the compact native structure. These data seem to point to a domain-wise folding mechanism based on hydrogen bonded secondary structures, akin to the backbone folding mechanism mentioned above. Thus for this mechanism, hydrogen bonding initiates folding and hydrophobic collapse promotes stabilization.

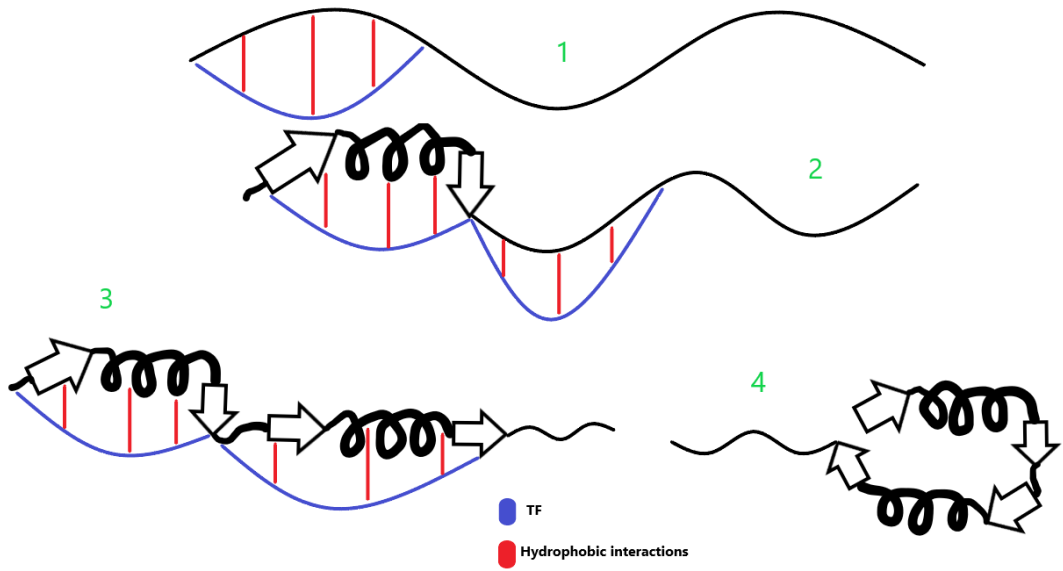

**FIG 3.** Domain Folding Theory based on first TF mechanism from *in vitro* studies of TF.

**The Domain-Microdomain Folding Theory:** The domain-microdomain folding theory is supported by a recent *in vivo* TF study by Oh et al, which found that TF firstly engages ~135 residues of nascent polypeptide chains, in the same way as in the first mechanism, after which it engages the chains at regular intervals of ~45 residues in a bind and release fashion.[14] **(Fig 4)** This TF mechanism also points to a folding mechanism where a domain is the first folded structure, after which the rest of the protein chain folds in a microdomain-wise fashion. As in the domain theory, folding is initiated by hydrogen bonding and hydrophobic collapse completes it. This folding mechanism may be validated by stability concerns. i.e early in folding, a domain which is much more stable than a microdomain, is folded first as a base folded structure, upon which subsequent "marginally stable" microdomains are added. This ensures that the folding process begins on a stable foundational domain structure, on which the complete protein structure is built, predominantly through fast folding microdomains.

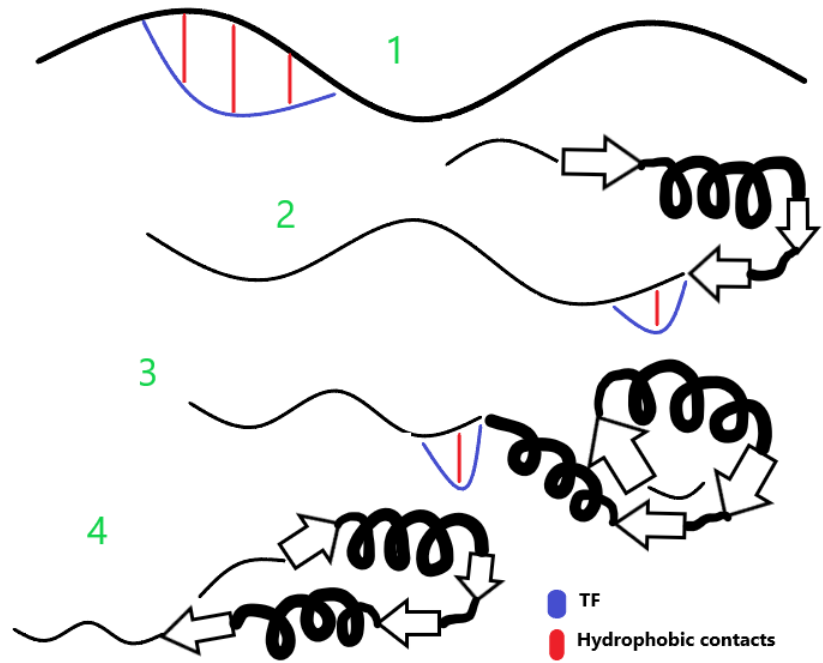

**FIG 4**. Domain-Microdomain Folding Theory based on second TF mechanism from *in vivo* studies of TF.

**DISCUSSION:** How do we know which of these theories is closer to the truth? Both theories are strongly supported by the fact that local interactions (hydrogen bonded secondary structures) are preferred to non-local ones early in folding. In other words, early in folding, there is little compromise on large degrees of freedom; non-local interactions limit the conformational freedom of polypeptide chains. Indeed, simple exact lattice model studies demonstrate that open conformations are closer to the native state than compact ones.[18] Both theories also simplify the folding problem as proposed in the diffusion-collision model. However, the domain-microdomain theory further simplifies the folding problem, by breaking it into much more smaller pieces than the domain theory. In a sense, the domain-microdomain theory is a mesh of the backbone hydrogen bonding theory and the diffusion-collision model. The domain-microdomain theory is almost the same as the diffusion-collision model. However, there is a slight difference between the two theories; in the domain-microdomain theory, a domain is the first folded structure unlike the microdomain in the diffusion-collision model. To what extent can the folding problem be broken down to smaller parts? At the moment, the limit is microdomain secondary structures, unless simpler folded structures are discovered. It is worth noting that the first TF mechanism on which the domain theory is based was identified via an *in vitro* TF study, and may not represent how TF actually helps proteins to fold in the cellular environment. In contrast, the *in vivo* TF study on which the domain-microdomain theory is based may show the true TF mechanism by which proteins fold. These considerations give much more credence to the domain-microdomain theory than the domain theory. Additionally, in the *in vivo* TF study, TF was found to predominantly interact with outer-membrane proteins, and thus the folding theories which have been described here may not apply to all proteins as a general folding mechanism.

**CONCLUSION:** The protein folding mechanism has not been completely understood, although a number of ideas have been proposed. However, it is understood that the mechanism follows a divide and conquer approach in order to solve the conformational search problem. Although the theories that I have discussed here are based on the mechanism of only one chaperone, they still shed some light on the protein folding mechanism. It would be interesting to look at other chaperone mechanisms and see whether similar folding theories might hold.

**ACKNOWLEDGEMENTS:** I thank Dr. George Rose and Dr. Ken Dill for useful criticisms, comments and suggestions.